\documentclass[%
 reprint,
 amsmath,amssymb,
 aps,
 prx,
floatfix,
]{revtex4-2}

\usepackage{graphicx}
\usepackage{dcolumn}
\usepackage{bm}
\usepackage{chemformula}
\usepackage{bbold}
\usepackage{dsfont}
\usepackage{xcolor}
\definecolor{costumgreen}{rgb}{0, 0.91, 0.047} 
\definecolor{costumred}{rgb}{1, 0, 0.0625}
\definecolor{costumblue}{rgb}{0, 0.238, 0.918}

\usepackage{xr-hyper}
\usepackage[linesnumbered,ruled,vlined]{algorithm2e}
\usepackage{algpseudocode}
\usepackage{amssymb}
\usepackage{xspace}
\usepackage{caption}
\usepackage{subcaption}
\usepackage{hyperref}

\usepackage{amsthm}

\newcommand{\integ}[4]{\int_{#1}^{#2}#3\text{d}#4}
\newcommand{\graph}{\mathcal{G}}
\newcommand{\tder}{\frac{\text{d}}{\text{d}t}}

\begin{document}

\preprint{???}

\title{Mutual linearity in semi-Markov networks}

\author{Julian B. Voits\textsuperscript{1,2}}
\author{Ulrich S. Schwarz\textsuperscript{1,2}}%
 \email{Corresponding author: schwarz@thphys.uni-heidelberg.de}
\affiliation{%
\textsuperscript{1}Institute for Theoretical Physics, Heidelberg University, Germany\\ \textsuperscript{2}BioQuant-Center for Quantitative Biology, Heidelberg University, Germany
}%




\date{\today}

\begin{abstract}
Despite their microscopic complexity, stochastic networks
show a surprising degree of universal, parameter-independent properties. One example is mutual linearity, which has recently been characterized as a generic, pairwise-linear relation between observables of Markov processes under local perturbations, valid for both stationary dynamics and for transient dynamics in Laplace space. Here we show that mutual linearity arises if a certain global affine relation is satisfied. This allows us to demonstrate that mutual linearity also holds for a broad class of local perturbations in semi-Markov networks, which, unlike Markov networks, have memory in terms of the dwell time in the current state.
However, exceptions exist, meaning that mutual linearity does not extend to all possible perturbations of semi-Markov networks. Strikingly, this restriction disappears for first-passage times. Thus, semi-Markov networks exhibit a fundamental asymmetry between occupation and first-passage statistics, different from Markov networks. Finally, we show that our results are directly applicable to stochastic resetting in semi-Markov networks, where the Laplace-space relations acquire a physical interpretation in terms of stationary occupations and the probability to reach a target before reset. This provides a natural setting in which the predicted affine relations can be probed. 
\end{abstract}

\maketitle

\textit{Introduction--} Stochastic networks constitute a fundamental class of models for complex systems, with applications ranging from physics, chemistry and biology~\cite{schnakenberg1976network,gillespie1992rigorous,gardiner2004handbook,van2004stochastic,honerkamp1996stochastic,bressloff2014stochastic,tang2021topology} to epidemiology~\cite{keeling2008modeling,kiss2017mathematics} and finance~\cite{d2006homogeneous,bardoscia2021physics,selvamuthu2025introduction}. Physical quantities of interest for these networks include the occupation probabilities of the states in the network and observables derived from them, e.g.\ entropy production and work, which are the central concepts in stochastic thermodynamics~\cite{seifert2012stochastic,seifert2025stochastic}.
Another important readout of stochastic networks is 
first-passage times (FPTs), which describe the time it takes for a stochastic process to reach a certain target state~\cite{redner2001guide, metzler2014first}. In biological systems, FPTs often mark the moment when a critical decision is made, e.g.\ for cell division or the immune response~\cite{chou2014first,iyer2016first,bel2010simplicity,wong2026first}.

Many real-world networks are characterized by a multitude of microscopic parameters whose values are often hard to determine, making non-model-specific results regarding the properties of such systems particularly desirable~\cite{gutenkunst2007universally, voits2025generic, voits2026emergence}.  
One example of such a universal result is mutual linearity in Markov networks. Originally, it was introduced as an affine relation between currents under the perturbation of a single transition rate, valid for both stationary dynamics and for non-stationary dynamics in Laplace space~\cite{harunari2024mutual, polettini2025coplanarity, khodabandehlou2025affine}. Recently, mutual linearity was shown to hold also for occupation probabilities in the steady state~\cite{floyd2025limits,bebon2026mutual} and in the transient regime in Laplace space~\cite{zheng2026mutual,voits2026graph}. Furthermore, it extends to first-passage time densities~\cite{voits2026graph}. An analogous relation also holds for continuous Markov processes~\cite{zheng2026mutual2} and activities in stationary chemical reaction networks~\cite{harunari2026mutual}, placing mutual linearity within the broader context of exact non-equilibrium response relations~\cite{baiesi2009fluctuations,altaner2016fluctuation,aslyamov2024nonequilibrium,aslyamov2025nonequilibrium,aslyamov2026dynamical}.
However, the results on mutual linearity have so far remained restricted to the case of memory-less stochastic dynamics, and an extension of this property to systems with memory is still lacking.  

Here, we extend mutual linearity under local perturbations to the broader class of semi-Markov networks~\cite{cinlar1975introduction,limnios2012semi}, which often arise from coarse-graining and partially accessible information on the states of the network~\cite{martinez2019inferring,van2022thermodynamic}. We show that
mutual linearity arises from a general global affine relation, valid for the Laplace transforms of the occupation probabilities under additional assumptions on the shape of the local perturbation.
We find that this relation holds for many, but not all perturbations of semi-Markov networks.
Remarkably, for first-passage time densities, it remains valid for arbitrary local perturbations, revealing a structural asymmetry between occupation probabilities and FPT densities for semi-Markov networks that does not exist for Markov networks.

As an illustrative example, we show that our results are directly applicable to the problem of stochastic resetting in semi-Markov networks, which describes the effect of introducing a random re-initialization of the stochastic process, leading to a non-equilibrium steady state and shorter first-passage times in target searches at optimal resetting rates~\cite{evans2020stochastic, pal2017first}. Stochastic resetting was originally studied for continuous processes such as Brownian motion~\cite{evans2011diffusion, evans2011diffusion2}, including experimental realizations~\cite{tal2020experimental} and later extended to other stochastic processes such as random walks on complex networks~\cite{avrachenkov2018hitting,riascos2020random}. It is
also an important element of kinetic proofreading networks~\cite{mckeithan1995kinetic,bel2010simplicity}.
These results suggest a possible entry-point for experimental verification of mutual linearity for semi-Markov processes.

\textit{Global affine response--} A semi-Markov network on $N$ states is defined on a graph $\graph=(V(\graph),E(\graph))$ with vertex set $V(\graph)=\{1,...,N\}$ and edge set $E(\graph)\subseteq\{(i,j)\in V(\graph)\times V(\graph)~|~i\neq j\}$, with transition laws of the shape
\begin{align}
    \big(Q_{ij}(t)\big)_{(i,j)\in E(\graph)}=\big(p_{ij}\psi_{ij}(t)\big)_{(i,j)\in E(\graph)},
\end{align}
where $p_{ij}>0$ is the probability that a process in state $i$ transitions next to state $j$, and $\psi_{ij}(t)$ denotes the probability density of the waiting time conditional on this transition. The distribution of the sojourn time in state $i$ is given by $Q_i(t):=\sum_lQ_{il}(t)$. The time-evolution of the occupation probabilities $\mathbf{p}(t)=\big(p_i(t)\big)_{i\in\{1,...,N\}}$ of a semi-Markov process $X_t$ started at $t=0$ on the network is governed by a generalized master equation:
\begin{align}
    \tder p_i(t)&=\sum_{j=1}^N\integ{0}{t}{W_{ij}(t-t')p_j(t')}{t'},\label{eq:generalizedME}
\end{align}
where
\begin{align}
    W_{ij}(t):=\begin{cases}
        ~~~~K_{ji}(t)&\text{if }(j,i)\in E(\graph)\\
        -\sum_{l}K_{il}(t)&\text{if }i=j\\
        \qquad0 &\text{else}.
    \end{cases}
\end{align}
The memory kernels $K_{ij}(t)$ can be expressed in Laplace space as
\begin{align}
    \tilde{K}_{ij}(s)=\frac{s\tilde{Q}_{ij}(s)}{1-\tilde{Q}_{i}(s)},
\end{align}
where the tilde denotes the Laplace transform. In Laplace space, Eq.~(\ref{eq:generalizedME}) takes the form of an algebraic equation,
\begin{align}
    \big(s\mathbb{1}-\tilde{W}(s)\big)\tilde{\mathbf{p}}(s)=\mathbf{p}_0,\label{eq:gME_Laplace_space}
\end{align}
under the initial condition $\mathbf{p}(0)=\mathbf{p}_0$.
In the Markovian case, the memory kernels are the transition rates $\tilde{K}_{ij}(s)=k_{ij}=\text{const}$ and the
transition densities $\psi_{ij}(t)$ are (memory-less) exponential 
functions. 

Now consider a perturbation of the system that can be
parametrized by a scalar $\lambda$. We first note
that when the response of $\tilde{\mathbf{p}}(s)$ is of the shape
\begin{align}
    \tilde{\mathbf{p}}(s,\lambda)=\tilde{\mathbf{a}}(s)+c(s,\lambda)\tilde{\mathbf{b}}(s),\label{eq:global_affine_relation}
\end{align}
then we obtain the mutual linearity relation immediately by
combining the derivatives in components $n$ and $m$,
\begin{align}
    \partial_\lambda\tilde{p}_n(s,\lambda)&=\partial_\lambda c(s,\lambda)\tilde{b}_n(s)=\frac{\tilde{b}_n(s)}{\tilde{b}_m(s)}\partial_\lambda\tilde{p}_m(s,\lambda)\\ \Rightarrow \tilde{p}_n(s,\lambda)&=\tilde{q}_{nm}(s)+\tilde{\chi}_{nm}(s)\tilde{p}_m(s,\lambda),\label{eq:mutual_linearity_occupation_probabilities}
\end{align}
where the susceptibility is defined as $\tilde{\chi}_{nm}(s):=\frac{\tilde{b}_n(s)}{\tilde{b}_m(s)}$. The following two relations are immediate consequences of the definition:
\begin{align}
    \tilde{\chi}_{nm}(s)=\tilde{\chi}_{mn}(s)^{-1},\qquad\tilde{\chi}_{nm}(s)=\tilde{\chi}_{nl}(s)\tilde{\chi}_{lm}(s),
\end{align}
in accordance with the previously identified properties of the susceptibilities in the Markovian case~\cite{bebon2026mutual,voits2026graph}.
In the supplement, we show that a combinatorial expression can be given for $\tilde{\chi}_{nm}(s)$, using the graph theoretic framework for semi-Markov networks developed in \cite{voits2026framework}. 

If the matrix $s\mathbb{1}-\tilde{W}(s,\lambda)$ can be decomposed as
\begin{align}
    s\mathbb{1}-\tilde{W}(s,\lambda)=\tilde{A}(s)+\tilde{\mathbf{u}}(s)\tilde{\mathbf{v}}(s,\lambda)^T,\label{eq:matrix_rank_one_decomposition}
\end{align}
where $\tilde{A}(s)$ is invertible, Eq.~(\ref{eq:gME_Laplace_space}) can be solved using the Sherman-Morrison inversion formula~\cite{sherman1950adjustment},
\begin{align}
    \tilde{\mathbf{p}}(s,\lambda)&=\big(s\mathbb{1}-\tilde{W}(s,\lambda)\big)^{-1}\mathbf{p}_0\\ &=\tilde{A}(s)^{-1}\mathbf{p}_0-c(s,\lambda)\tilde{A}(s)^{-1}\tilde{\mathbf{u}}(s),
\end{align}
where 
\begin{align}
    c(s,\lambda):= \frac{\tilde{\mathbf{v}}(s,\lambda)^T\tilde{A}(s)^{-1}\mathbf{p}_0}{1+\tilde{\mathbf{v}}(s,\lambda)^T\tilde{A}(s)^{-1}\tilde{\mathbf{u}}(s)}.
\end{align}
With the identification $\tilde{\mathbf{a}}(s)=\tilde{A}(s)^{-1}\mathbf{p}_0$ and $\tilde{\mathbf{b}}(s)=-\tilde{A}(s)^{-1}\tilde{\mathbf{u}}(s)$, we see that exactly the global affine relation from Eq.~(\ref{eq:global_affine_relation}) arises.

To identify the class of local perturbations that imply a global affine relation of the occupation probabilities in Laplace space as in Eq.~(\ref{eq:global_affine_relation}) and therefore mutual linearity, we note that any variation of the outgoing transition laws of a state $n\in V(\graph)$ along a one-dimensional affine subspace for a fixed $s$,
\begin{align}
    \tilde{Q}_{nl}(s)\to \tilde{Q}_{nl}(s)+\alpha(s,\lambda) \tilde{\theta}_{nl}(s)\label{eq:perturbation_Q},
\end{align}
gives rise to a decomposition of $s\mathbb{1}-\tilde{W}(s,\lambda)$ as in Eq.~(\ref{eq:matrix_rank_one_decomposition}). Details and explicit expressions for $\tilde{A}(s)$, $\tilde{\mathbf{u}}(s)$ and $\tilde{\mathbf{v}}(s,\lambda)$ are given in appendix A.

A simple realization of a perturbation as in Eq.~(\ref{eq:perturbation_Q}) is a variation of a single transition law, $\tilde{\theta}_{nl}(s)=\delta_{nm}$. In particular, an arbitrary perturbation to a state with a single outgoing transition will imply a linearity relation as in Eq.~(\ref{eq:global_affine_relation}). However, for multiple outgoing edges, the constraint that $\sum_lp_{nl}=1$ means that a perturbation of a single transition law is only possible at fixed transition probabilities, i.e., it may only perturb $\psi_{ij}(t)\to \psi_{ij}(t,\lambda)$.

For a variation of the exit probabilities, Eq.~(\ref{eq:perturbation_Q}) 
arises for linear perturbations of the shape
\begin{align}
    p_{nl}\to p_{nl}+\alpha(\lambda) q_{l}\in [0,1],\label{eq:perturbations_exit_probabilities}
\end{align}
for one $n\in V(\graph)$.
Since the total exit probability must sum up to one, this requires $\sum_lq_{l}=0$.
Important examples of variations of this shape include any redistribution between two exit probabilities to states $m_1,m_2\in V(\graph)$ while keeping the other ones fixed,
\begin{align}
    q_l=\delta_{lm_1}-\delta_{lm_2},
\end{align}
and any perturbation that proportionally redistributes the probability among the exit channels, i.e.,
\begin{align}
    q_l= (1-p_{nm})\delta_{lm}-p_{nl}(1-\delta_{lm}).
\end{align}

An important subclass of semi-Markov networks assumes direction-time independence, i.e.,
\begin{align}
    Q_{ij}(t)=p_{ij}\psi_i(t).
\end{align}
In particular, $Q_{i}(t)=\psi_i(t)\sum_jp_{ij}=\psi_i(t)$. 
Now consider a local perturbation $\lambda$ of the shape 
\begin{align}
    \tilde{Q}_{nl}(s)\to \tilde{Q}_{nl}(s,\lambda)=(r_l(s)+f(s,\lambda)\tilde{q}_l(s))\tilde{\psi}_n(s,\lambda)\label{eq_perturbation_direction_time_independence}
\end{align}
for one $n\in V(\graph)$. If $f(s,\lambda)\tilde{\psi}_n(s,\lambda)=a(s)+b(s)\tilde{\psi}_n(s,\lambda)$, $s\mathbb{1}-\tilde{W}(s,\lambda)$ can be written in the shape of Eq.~(\ref{eq:perturbation_Q}) with $\alpha(s,\lambda)=\tilde{\psi}_n(s,\lambda)$ and $\tilde{\theta}_{nl}(s)=b(s)\tilde{q}_l(s)$.
In particular, Markov networks fall into this class as their transition laws are given by
\begin{align}
    \tilde{Q}_{nl}(s)=\frac{k_{nl}}{s+\sum_j k_{nj}}=\frac{k_{nl}}{\sum_j k_{nj}}\frac{\sum_j k_{nj}}{s+\sum_j k_{nj}}.
\end{align}
If a single exit rate is perturbed, say $k_{nm}\to \lambda$, then we may write
\begin{align}
    \tilde{Q}_{nm}(s,\lambda)&=\Big(1-\frac{\sum_{j\neq m} k_{nj}}{\sum_{j\neq m} k_{nj}+\lambda}\Big)\frac{\sum_{j\neq m} k_{nj}+\lambda}{s+\sum_{j\neq m} k_{nj}+\lambda}
\end{align}
so we see that such a perturbation is of the shape of Eq.~(\ref{eq_perturbation_direction_time_independence}) with $r_l(s)=\delta_{ml}$, $f(s,\lambda)=(\sum_{j\neq m} k_{nj}+\lambda)^{-1}$ and 
\begin{align}
    q_l=k_{nl}(1-\delta_{ml})-\sum_{j\neq m} k_{nj}\delta_{ml}.
\end{align}
Hence, the mutual linearity for Markov networks relation previously shown in \cite{zheng2026mutual, voits2026graph} follows as a special case. However, the class of variations that will give rise to mutual linearity in general semi-Markov networks is not exhaustive
and we will describe exceptions below.

\textit{First-passage times--} We consider the first-passage time (FPT) $\tau$ to some state of the network, which we choose to be $N$, without loss of generality. Moreover, since only the dynamics until $N$ is reached for the first time is relevant for the FPT, the state $N$ can be assumed to be absorbing, i.e., $(N,l)\notin E(\graph)$ for all $l\in V(\graph)$. Then, the process remains in $N$ once this state is reached and we have
\begin{align}
    \mathbb{P}(\tau\leq t)=p_N(t),
\end{align}
so $p_N(t)$ yields the cumulative distribution function and its density is obtained by differentiation,
\begin{align}
    f(t)=\frac{\text{d}}{\text{d}t}\mathbb{P}(\tau\leq t)=\dot{p}_N(t).
\end{align}
Using Eq.~(\ref{eq:generalizedME}) and assuming no initial occupation of the target state, $p_N(t=0)=0$, we obtain an expression for the FPT density in Laplace space,
\begin{align}
    \tilde{f}(s)&=s\tilde{p}_N(s)
    \\ &= s\mathbf{p}_0^T(s\mathbb{1}-\tilde{W}(s)^T)^{-1}\hat{e}_N,\label{eq:universal_FPT_relation}
\end{align}
where $\hat{e}_N$ denotes the unit vector in $N$-direction. Writing $f_i(t):=f(t~|~p_{0,j}=\delta_{ij})$, we have
\begin{align}
    \tilde{\mathbf{f}}(s)= s(s\mathbb{1}-\tilde{W}(s)^T)^{-1}\hat{e}_N,
\end{align}
which is analogous to Eq.~(\ref{eq:gME_Laplace_space}) but with the key difference that the transpose of $\tilde{W}(s)$ enters in the description for the first-passage time. As a result, any local perturbation,
\begin{align}
    \tilde{Q}_{nl}(s)\to \tilde{Q}_{nl}(s,\lambda),
\end{align}
for one $n\in V(\graph)$, can be expressed as 
\begin{align}
    s\mathbb{1}-\tilde{W}(s,\lambda)^T=\tilde{A}(s)+\hat{e}_{n}\tilde{\mathbf{v}}(s,\lambda)^T,
\end{align}
where 
\begin{align}
    \tilde{A}_{ij}(s):=&(s\mathbb{1}-\tilde{W}(s,\lambda))_{ji}(1-\delta_{in})+s\delta_{ij}\delta_{in}, \label{eq:A_matrix_FPT}\\ \tilde{v}_i(s,\lambda)=&\sum_l\tilde{K}_{nl}(s,\lambda)\delta_{in}-\tilde{K}_{ni}(s,\lambda)(1-\delta_{in}).
\end{align}
Hence, $\tilde{\mathbf{f}}(s)$ can be written in the shape
\begin{align}
     \tilde{\mathbf{f}}(s)&=s\tilde{A}(s)^{-1}\hat{e}_N-sd(s,\lambda)\tilde{A}(s)^{-1}\hat{e}_n,
\end{align}
where 
\begin{align}
    d(s,\lambda):= \frac{\tilde{\mathbf{v}}(s,\lambda)^T\tilde{A}(s)^{-1}\hat{e}_N}{1+\tilde{\mathbf{v}}(s,\lambda)^T\tilde{A}(s)^{-1}\hat{e}_n},
\end{align}
implying that a mutual linearity relation holds for the Laplace transforms of the FPT densities starting at different initial states,
\begin{align}
    \tilde{f}_i(s,\lambda)=\tilde{g}_{ij}(s)+\tilde{\xi}_{ij}(s)\tilde{f}_j(s,\lambda)\label{eq:mutual_linearity_FPT}
\end{align}
for arbitrary local perturbations, in contrast to the occupation probabilities.

\begin{figure*}[t]
    \centering
    \includegraphics[width=1\linewidth]{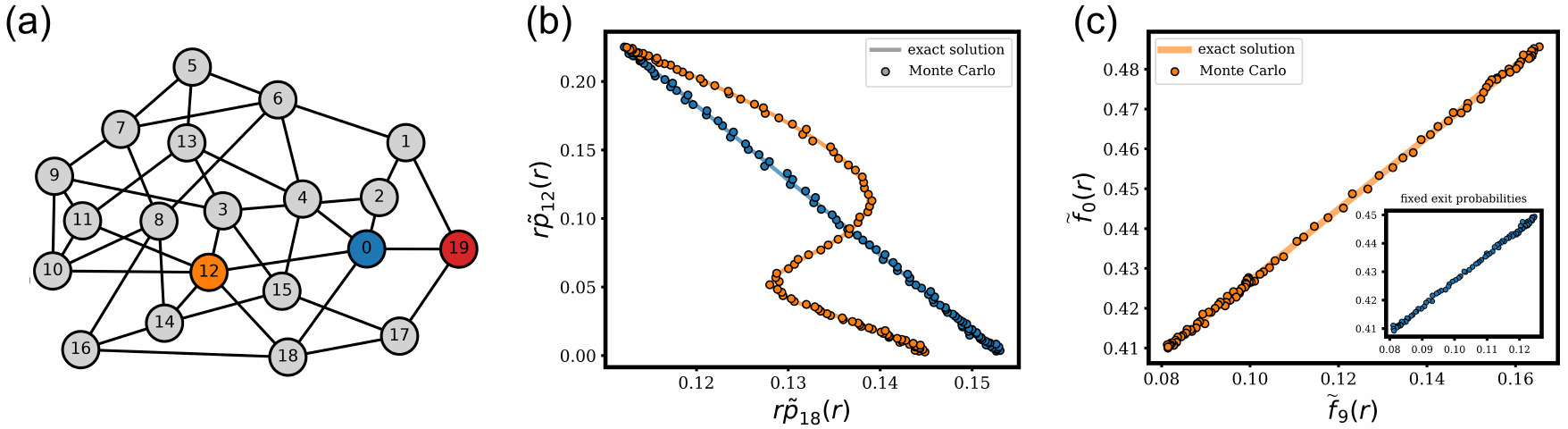}
    \caption{Illustration of the linearity relation for a search process with resetting on a random network. (a) Randomly generated small-world network with $20$ vertices. The process is initialized in and reset to state $0$ (blue), state $19$ (red) is the target and state $12$ (orange) the perturbed state. (b) Exact solution and Monte Carlo data points ($10^6$ runs each) for the relation between two stationary occupation probabilities for the network with resetting. For fixed exit probabilities, the two occupation probabilities are linearly related (blue), while for an additional non-linear perturbation, the relation need not be linear (orange). (c) The probabilities to reach the target state for different initial states are mutually linear also for arbitrary perturbations of the exit probabilities. The inset shows the relation for fixed exit probabilities for comparison.}
    \label{fig:resetting_illstration}
\end{figure*}

\textit{Stochastic resetting--} To illustrate our results on mutual linearity for semi-Markov networks, including cases in which it does not hold, we now turn to stochastic resetting. As an example, consider a random walk-like process on a finite graph $\graph$. Stochastic resetting means restarting the process at a certain vertex $n\in V(\graph)$. In the simplest case, the resetting of the process occurs at a constant rate $r$~\cite{evans2011diffusion}, e.g., a constant dissociation rate in kinetic proofreading~\cite{bel2010simplicity}, but extensions to non-Poissonian resetting (i.e., with memory) have also been considered~\cite{evans2020stochastic}. It has been studied for discrete-time and continuous-time Markov processes on complex networks~\cite{riascos2020random}. Here, we consider the more general setting of  state-dependent, generally non-exponential sojourn time distributions and arbitrary exit probabilities, i.e., $\tilde{Q}_{ij}(s)=p_{ij}\tilde{\psi}_i(s)$. As a specification for semi-Markov networks, we understand a reset as also restarting the waiting time from zero.
The occupation probability $p_i(r,t)$ of state $i$ at time $t$ and reset rate $r$ for a process initialized in the reset state can be expressed as 
\begin{align}
    p_i(r,t)= e^{-rt}p_i(0,t)+r\integ{0}{t}{p_i(0,t') e^{-rt'}}{t'},\label{eq:reset_probability_decomposition}
\end{align}
where the first term describes the case that the process has not been reset until $t$ and the second term integrates over the time since the last reset. Taking the long-time limit of Eq.~(\ref{eq:reset_probability_decomposition}) yields    
\begin{align}
    \lim_{t\to\infty}p_i(r,t)=r\integ{0}{\infty}{p_i(0,t') e^{-rt'}}{t'}=r\tilde{p}_i(0,r),
\end{align}
so the Laplace transform carries the direct physical meaning of the steady state of the process under resetting.

Moreover, the probability of reaching the target within a reset interval can be expressed as
\begin{align}
    \integ{0}{\infty}{re^{-rt}\integ{0}{t}{f_n(t')}{t'}}{t}=\integ{0}{\infty}{f_n(t)e^{-rt}}{t}=\tilde{f}_n(r),
\end{align}
using integration by parts and the fact that $e^{-rt}\integ{0}{t}{f_n(t')}{t'}$ vanishes at the boundaries. Thus, the Laplace transform of the FPT density can be interpreted to be exactly this probability.

Now consider a perturbation of one state $m\in V(\graph)$, such that $p_{mi}\to p_{mi}(\lambda)$ and $\psi_{m}(t)\to \psi_{m}(\lambda,t)$. Then, the probabilities that the target is reached before a reset for two different initial states $n_1,n_2\in V(\graph)$ are linearly related according to Eq.~(\ref{eq:mutual_linearity_FPT}) for an arbitrary perturbation of the state $m$. The stationary probabilities $r\tilde{p}_i(r)$ of the process under resetting are also linearly related  if the perturbation is of the shape of Eq.~(\ref{eq_perturbation_direction_time_independence}). We demonstrate this for a random small-world network~\cite{watts1998collective,avrachenkov2018hitting, riascos2020random} with gamma-distributed sojourn times,
\begin{align}
    \psi_n(t)=\frac{t^{k_n-1}}{\Gamma(k_n)\theta_n^{k_n}}e^{-\frac{t}{\theta_n}},
\end{align}
and transition probabilities $p_{ij}$ drawn from a uniform distribution on the simplex $\sum_{j:(i,j)\in E(\graph)}p_{ij}=1$. We choose a perturbed state $m\in V(\graph)$ with a sojourn-time density parametrized as
\begin{align}
    \psi_m(t,\lambda )=\frac{t^{\lambda-1}}{\Gamma(\lambda)e^{\lambda^3}}e^{-te^{-\lambda^2}},
\end{align}
without perturbation of the exit probabilities in the first case and with
\begin{align}
    p_{ml_j}=\frac{\lambda^{j}}{\sum_{k=1}^n\lambda^{k}}\label{eq:non_linear_perturbation_probabilities}
\end{align}
for the vertices $l_1<l_2<...<l_n$ adjacent to $m$ in the second case. The shape of Eq.~(\ref{eq:non_linear_perturbation_probabilities}) serves as an example of a non-linear local variation of the transition probabilities, which does not guarantee a linear relation between the occupation probabilities.

Fig.~\ref{fig:resetting_illstration} shows the results of our simulations, which confirm the predicted behaviour. The randomly generated small-world network with $20$ vertices is shown in Fig.~\ref{fig:resetting_illstration}(a). The stochastic process is initialized in state $0$ and for the first-passage time, state $19$ was chosen as the target. The perturbed state was chosen among the states with the most outgoing edges, excluding the initial and target states. 
Fig.~\ref{fig:resetting_illstration}(b) and Fig.~\ref{fig:resetting_illstration}(c) show the stationary occupation probabilities of two states as a function of one another and the probability of reaching the target state before the process is reset for two different initial states, respectively. The diagrams show the results for fixed transition  probabilities (blue) and for a non-linear variation of them as in Eq.~(\ref{eq:non_linear_perturbation_probabilities}) (orange). The estimates of the Monte Carlo simulations on the network (circles) correspond to $10^6$ runs each and are in excellent agreement with the exact solutions $r\tilde{p}_i(r)$ and $\tilde{f}_i(r)$ (solid lines) obtained by using the Laplace space representation of the gamma-distributed sojourn times, $\tilde{\psi}_i(s)=(1+\theta_is)^{-k_i}$ and evaluating Eqs.~(\ref{eq:gME_Laplace_space}) and (\ref{eq:universal_FPT_relation}) by numerically inverting $s\mathbb{1}-\tilde{W}(s)$. 
More details on the simulations are given in appendix B.

As predicted, the FPT densities in Laplace space, i.e., the probabilities to reach the target state before the reset, are mutually linear in both cases, illustrating that this property is robust for arbitrary local perturbations. In contrast, the occupation probabilities in Laplace space, which correspond to the stationary occupation probabilities of the process with resetting, are mutually linear for a broad but not exhaustive class of local perturbations.

\textit{Discussion--} In this work, we have shown that the property of mutual linearity previously identified for the occupation probabilities and first-passage time densities of Markov processes follows from a general global affine relation (Eq.~(\ref{eq:global_affine_relation})), which allows for a generalization of the treatment of mutual linearity to semi-Markov networks. We have identified conditions which give rise to mutual linearity between the occupation probabilities of semi-Markov networks in Laplace space. These include Markov networks as a special case, recovering the known relations from previous work~\cite{zheng2026mutual,voits2026graph}. However, mutual linearity does not generally hold under arbitrary local perturbations in semi-Markov networks. Surprisingly, the relation remains valid for FPT densities in Laplace space for local perturbations of any shape, revealing an asymmetry not present
for Markov networks. We applied our results to stochastic resetting on semi-Markov networks, where the expressions in Laplace space have a direct interpretation in terms of physically observable quantities. In the future, this may lead to an experimental validation of mutual linearity without the constraints of stationarity and the Markov property of the underlying process.

New non-model-specific results on semi-Markov networks as
reported here are very timely in the light of the growing use of coarse-grained stochastic dynamics in recent work on thermodynamic inference from waiting-time statistics~\cite{van2022thermodynamic, ertel2024estimator, van2025thermodynamic}.  In the future, identifying universal constraints like the affine relation derived here may facilitate inference in stochastic networks with memory.
An interesting application of our results for the resetting might be 
kinetic proofreading with memory, e.g.\ for T-cell receptor signaling \cite{biswas2026proofreadingimprovesspeedaccuracy}.

\textit{Acknowledgments--} JBV thanks the German Academic Scholarship Foundation (Studienstiftung des deutschen Volkes) for support. 
We also acknowledge support by the Max Planck School Matter to Life funded by the Dieter Schwarz Foundation and the Max Planck Society.

\textit{Data availability--} The code used to generate the data that support the findings of this article is openly available~\cite{Github2026}.


\vspace{2cm}

\bibliography{apssamp}
\newpage 

\appendix
\section*{Endmatter}

\textit{Appendix A: Affine perturbations of the local transition laws--} In the main text, we considered perturbations of the shape
\begin{align}
    \tilde{Q}_{nl}(s)\to \tilde{Q}_{nl}^{(p)}(s)+\alpha(s,\lambda) \tilde{\theta}_{nl}(s),
\end{align}
allowing for the rank-one decomposition of $s\mathbb{1}-\tilde{W}(s,\lambda)$ that implies mutual linearity.
To see that this is indeed the case, we here give the explicit construction of $\tilde{A}(s)$, $\tilde{u}(s)$ and $\tilde{v}(s)$ as in Eq.~(\ref{eq:matrix_rank_one_decomposition}). 
First, we need to distinguish the two cases $\tilde{\theta}_{n}(s):=\sum_l \tilde{\theta}_{nl}(s)=0$ and $\tilde{\theta}_{n}(s)\neq 0$. In the former case, we have
\begin{align}
    \tilde{K}_{nl}(s,\lambda)&=s\frac{\tilde{Q}^{(p)}_{nl}(s)}{1-\tilde{Q}^{(p)}_n(s)}+\alpha(s,\lambda) \frac{s\tilde{\theta}_{nl}(s)}{1-\tilde{Q}^{(p)}_n(s)}.\label{eq:K_perturbed_1}
\end{align}

For $\tilde{\theta}_{n}(s)\neq 0$, the transition kernels can be written as
\begin{align}
    \tilde{K}_{nl}(s,\lambda)&=s\frac{\tilde{Q}^{(p)}_{nl}(s)+\alpha(s,\lambda) \tilde{\theta}_{nl}(s)}{1-(\tilde{Q}^{(p)}_n(s)+\alpha(s,\lambda) \tilde{\theta}_{n}(s))}\\ &=s\big(\frac{\tilde{Q}_{nl}^{(p)}(s)+(1-\tilde{Q}^{(p)}_n(s))\frac{\tilde{\theta}_{nl}(s)}{\tilde{\theta}_n(s)}}{1-(\tilde{Q}^{(p)}_n(s)+\alpha(s,\lambda) \tilde{\theta}_{n}(s))}-\frac{\tilde{\theta}_{nl}(s)}{\theta_{n}(s)}\big).\label{eq:K_perturbed_2}
\end{align}
Thus, in both cases, the memory kernels can be written as
\begin{align}
    \tilde{K}_{nl}(s,\lambda)=a_{nl}(s)+\beta(s,\lambda)b_{nl}(s).
\end{align}
Choose any reference value $\lambda_0$, e.g., value of $\lambda$ corresponding to the unperturbed process. Defining
\begin{align}
    \tilde{u}_i(s)&:=-b_{ni}(s)(1-\delta_{in})+\sum_lb_{nl}(s)\delta_{in},\\ \tilde{v}_i(s,\lambda)&:=(\beta(s,\lambda)-\beta(s,\lambda_0))\delta_{in},\\ \tilde{A}(s)&:=s\mathbb{1}-\tilde{W}(s,\lambda_0),
\end{align}
we see that $s\mathbb{1}-\tilde{W}(s,\lambda)$ can indeed be written in the shape of Eq.~(\ref{eq:matrix_rank_one_decomposition}). Moreover, $\tilde{A}(s)$ is invertible for $s>0$, as shown in the supplement.

\textit{Appendix B: Additional details on the simulations--} The network was generated as a random connected Watts-Strogatz graph with $V(\graph)=\{0,...,19\}$, initial neighborhood size $4$ and rewiring probability $0.3$ as a model for a small-world network. The graph is undirected and we interpret every edge $\{i,j\}\in E(\graph)$ as transitions in both directions,
\begin{align}
    \{i,j\}\leftrightarrow (i,j), (j,i).
\end{align}

For every state $n\in V(\graph)$, a shape $k_n$ and a scale $\theta_n$ parameter are randomly drawn from uniform distributions $k_n\in (1.2, 3.5)$ and $\theta_n\in (0.4,1.2)$ and the transition probabilities for the outgoing edges $(n,j)\in E(\graph)$ were drawn uniformly on the simplex $\sum_jp_{nj}=1$.

The perturbed state was chosen among the states with the highest number of outgoing edges excluding the initial and the target state. The shape of the perturbation was as specified in the main text, and $\lambda$ was varied from $\lambda=0.1$ to $\lambda=2.1$.

The stochastic dynamics on the network can be simulated by drawing a sojourn time according to the gamma distribution of the current state $n\in V(\graph)$ and by comparing this to a random reset time sampled from an exponential distribution with parameter $r=0.1$. If the sojourn time was less than the reset time, the next state $j$ was chosen according to the probability distribution $(p_{nj})_{(n,j)\in E(\graph)}$. Otherwise, the process was reset to the initial state.

To obtain Monte Carlo estimates for the stationary occupation probabilities, we exploited the memoryless property of Poissonian resetting by repeatedly sampling a global reset time $t_{\text{reset}}\sim \text{Exp}{(r)}$, letting the reset state evolve until that time and recording the final state. This is equivalent but more efficient than estimating the stationary occupation probabilities from one long-time trajectory.

Similarly, the FPT densities were obtained from repeated trajectories with stopping times $t_{\text{reset}}\sim \text{Exp}{(r)}$ initialized in the indicated states, which were stopped once they reached the target state or at the time $t_{\text{reset}}$. The Laplace transforms of the FPT densities were then estimated from the success probability of reaching the target state.




\end{document}


\title{Supplemental Material for\\
'Mutual linearity in semi-Markov networks'}

\author{Julian B. Voits\textsuperscript{1}}
\author{Ulrich S. Schwarz\textsuperscript{1,2}}%
 \email{Corresponding author: schwarz@thphys.uni-heidelberg.de}
\affiliation{%
\textsuperscript{1}Institute for Theoretical Physics, Heidelberg University, Germany\\ \textsuperscript{2}BioQuant-Center for Quantitative Biology, Heidelberg University, Germany
}%
\date{\today}

\maketitle

\section*{Graph theoretic meaning of the mutual linearity coefficients}
\bigskip


In this supplement, we apply the All-Minors Matrix-Tree theorem to show that the matrices in the main text are indeed invertible and that explicit combinatorial expressions for the susceptibilities from the mutual linearity relations can be given within the graph theoretic framework~\cite{nam2025algebraic,voits2026graph, voits2026framework}.

Let $\graph=(V(\graph),E(\graph))$, where $V(\graph)=\{1,...,N+1\}$ be a directed graph with edge weights $(w_{ij})_{(i,j)\in E(\graph)}$. Its (out-degree) Laplacian matrix $L$ is defined as
\begin{align}
    L_{ij}:=\begin{cases}
        ~~-w_{ji}&\text{if }(j,i)\in E(\graph)\\ ~\sum_l w_{il}&\text{if } i=j\\ \qquad0&\text{else}.
    \end{cases}
\end{align}
The All-Minors Matrix-Tree theorem gives a combinatorial expression for the determinants of the minors of $L$~\cite{chaiken1982combinatorial}. Here, we only need two special cases, namely the determinants of the $N+1$-th principal minor $L_{\{N+1\},\{N+1\}}$ and the minor obtained from deleting the rows $i\neq N+1$ and $N+1$ and the columns $j\neq N+1$ and $N+1$, $L_{\{i,N+1\},\{j,N+1\}}$. Defining the weight of a subgraph $\mathcal{S}\subseteq \graph$ as
\begin{align}
    w(\mathcal{S}):=\prod_{(i,j)\in E(\graph)}w_{ij},
\end{align}
the determinant of $L_{\{N+1\},\{N+1\}}$ is given by the total weight of all spanning trees rooted in $N+1$,
\begin{align}
    \det(L_{\{N+1\},\{N+1\}})=\sum_{\mathcal{T}_{[N+1]}}w(\mathcal{T}),
\end{align}
and the determinant of $L_{\{i,N+1\},\{j,N+1\}}$ is given by the total weight of all spanning forests with roots $i$ and $N+1$ containing a $j\to i$ path,
\begin{align}
    \det(L_{\{i,N+1\},\{j,N+1\}})=(-1)^{i+j}\sum_{\mathcal{F}^{j\to i}_{[i,N+1]}}w(\mathcal{F}).\label{eq:determinant_forest_identity}
\end{align}

\begin{figure}
    \centering
    \includegraphics[width=0.4\linewidth]{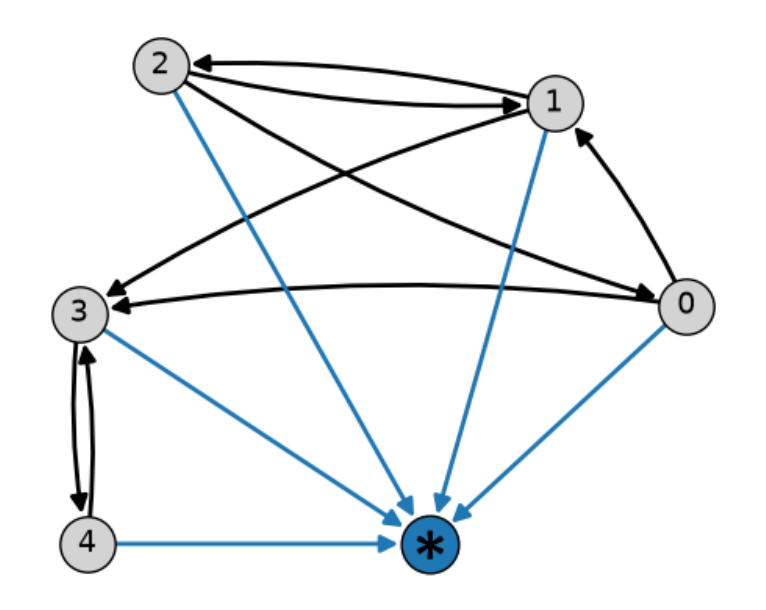}
    \caption{$s\mathbb{1}-\tilde{W}(s)$ can be interpreted as a principal minor of an weighted and directed graph, which is obtained from augmenting the original network by an additional vertex $\ast$ and adding a directed edge with weight $s$ leading from every vertex of the network to the additional state $\ast$. This allows us to apply the All-Minors Matrix-Tree theorem to give expressions for the elements of $\big(s\mathbb{1}-\tilde{W}(s)\big)^{-1}$ in terms of the spanning trees and spanning forests of the network.}
    \label{fig:augmented_graph}
\end{figure}

Observe that the matrix $s\mathbb{1}-\tilde{W}(s)$ is exactly of the shape of a principal minor $L_{\{N+1\},\{N+1\}}$ of a graph which is obtained by augmenting the original graph of the network $\graph$ by an additional vertex $\ast$ and connecting every other vertex via a directed edge (Fig.~\ref{fig:augmented_graph}), i.e.,
\begin{align}
    V(\graph^\ast)&=V(\graph)\cup\{\ast\},\\E(\graph^\ast)&=E(\graph)\cup\{(i,\ast)~|~i\in V(\graph)\},
\end{align}
and assigning the weight $w(i,j)=\tilde{K}_{ij}(s)$ to the original edges $(i,j)\in E(\graph)$ and the weight $w(i,\ast)=s$ to the additional edges. 
The matrix $s\mathbb{1}-\tilde{W}(s)$ can be interpreted as a principal minor of the Laplacian matrix of this graph obtained after removing the row and column corresponding to $\ast$. Thus, its determinant reads
\begin{align}
    \det{\big(s\mathbb{1}-\tilde{W}(s)\big)}=\sum_{\mathcal{T}^s_{[\ast]}}w(\mathcal{T})>0,
\end{align}
for $s>0$, where we use the notation $\mathcal{T}^s$ with the superscript $s$ to indicate that the trees are taken over the augmented graph. 

Moreover, we can use Eq.~(\ref{eq:determinant_forest_identity}) to compute the entries of the inverse of $s\mathbb{1}-\tilde{W}(s)$,
\begin{align}
    (s\mathbb{1}-\tilde{W}(s))_{ij}^{-1}&=\frac{(-1)^{i+j}\det(s\mathbb{1}-\tilde{W}(s))_{\{j\},\{i\}}}{\det(s\mathbb{1}-\tilde{W}(s))}\\ &=\frac{\sum_{\mathcal{F}^{s,j\to i}_{[i,\ast]}}w(\mathcal{F})}{\sum_{\mathcal{T}^s_{[\ast]}}w(\mathcal{T})}.
\end{align}
Applying the generalized master equation, this yields an explicit combinatorial expression for the occupation probabilities in Laplace space,
\begin{align}
    \tilde{p}_i(s)=\sum_{j=1}^N\frac{\sum_{\mathcal{F}^{s,j\to i}_{[i,\ast]}}w(\mathcal{F})}{\sum_{\mathcal{T}^s_{[\ast]}}w(\mathcal{T})}p_{0,j},\label{eq:transient_semi-Markov_Tree_theorem}
\end{align}
which can be seen as a generalization of the Markov Chain Tree theorem to transient occupation probabilities and semi-Markov networks.

Observe that due to the defining property of trees and forests to have no cycles, every term in the numerator and in the denominator of Eq.~(\ref{eq:transient_semi-Markov_Tree_theorem}) has at most one factor per vertex in $V(\graph)$.
Hence, for a local perturbation of some state $n\in V(\graph)$ of the form
\begin{align}
    \tilde{K}_{nl}(s,\lambda)=a_{nl}(s)+\beta_n(s,\lambda)b_{nl}(s),
\end{align}
the denominator in Eq.~(\ref{eq:transient_semi-Markov_Tree_theorem}) can be expressed as
\begin{align}
    \sum_{\mathcal{T}^s_{[\ast]}}w(\mathcal{T}) =&C(s)+\beta_n(s,\lambda)D(s),
\end{align}
where
\begin{align}
    C(s)&=s\sum_{\mathcal{F}^s_{[n,\ast]}}w(\mathcal{F})+\sum_la_{nl}(s)\sum_{\mathcal{F}^{s,l\to \ast}_{[n,\ast]}}w(\mathcal{F})\\ D(s)&=\sum_lb_{nl}(s)\sum_{\mathcal{F}^{s,l\to \ast}_{[n,\ast]}}w(\mathcal{F}).
\end{align}
Similarly, the numerator in Eq.~(\ref{eq:transient_semi-Markov_Tree_theorem}) can be expressed as
\begin{align}
    &\sum_{j=1}^Np_{0,j}\Big(s\sum_{\mathcal{F}^{s,j\to i}_{[i,n,\ast]}}w(\mathcal{F})+\sum_l\tilde{K}_{nl}(s,\lambda)\sum_{\mathcal{F}^{s,j\to i,l\not\to n}_{[i,n,\ast]}}w(\mathcal{F})\Big)\\ =&A_i(s)+\beta_n(s,\lambda)B_i(s),
\end{align}
where
\begin{align}
    A_i(s)&=\sum_{j=1}^Np_{0,j}\Big(s\sum_{\mathcal{F}^{s,j\to i}_{[i,n,\ast]}}w(\mathcal{F})+\sum_la_{nl}(s)\sum_{\mathcal{F}^{s,j\to i,l\not\to n}_{[i,n,\ast]}}w(\mathcal{F})\Big),
    \\ B_i(s)&=\sum_{j=1}^Np_{0,j}\sum_lb_{nl}(s)\Big(\sum_{\mathcal{F}^{s,j\to i,l\to i}_{[i,n,\ast]}}w(\mathcal{F})+\sum_{\mathcal{F}^{s,j\to i,l\to \ast}_{[i,n,\ast]}}w(\mathcal{F})\Big).
\end{align}

Eq.~(\ref{eq:transient_semi-Markov_Tree_theorem}) takes the shape
\begin{align}
    \tilde{p}_i(s,\lambda)&=\frac{A_i(s)+\beta_n(s,\lambda)B_i(s)}{C(s)+\beta_n(s,\lambda)D(s)}\\ &=\frac{A_i(s)-C(s)\frac{B_i(s)}{D(s)}}{C(s)+\beta_n(\lambda,s)D(s)}+\frac{B_i(s)}{D(s)}.
\end{align}
This shows that $\tilde{\mathbf{p}}(s,\lambda)$ is of the form of the global affine relation in the main text with the identifications
\begin{align}
    \tilde{a}_i(s)&=\frac{B_i(s)}{D(s)},\\ \tilde{b}_i(s)&=A_i(s)-C(s)\frac{B_i(s)}{D(s)},\\ c(s,\lambda)&=\frac{1}{C(s)+\beta_n(s,\lambda)D(s)}.
\end{align}
Hence, the susceptibility for the occupation probabilities reads
\begin{align}
    \tilde{\chi}_{ij}(s)&=\frac{\tilde{b}_i(s)}{\tilde{b}_j(s)}=\frac{A_i(s)D(s)-C(s)B_i(s)}{A_j(s)D(s)-C(s)B_j(s)},
\end{align}
with the graph theoretic meaning of $A_i(s), B_i(s), C(s)$ and $D(s)$ given above.

For the first-passage time densities, we obtain
\begin{align}
    f_i(s)&=s\tilde{p}_N(s~|~p_{0,j}=\delta_{ij})\\ &=s\frac{A^{(i)}_N(s)-C(s)\frac{B^{(i)}_N(s)}{D(s)}}{C(s)+\beta_n(\lambda,s)D(s)}+s\frac{B^{(i)}_N(s)}{D(s)},
\end{align}
where 
\begin{align}
    A^{(i)}_N(s)&=\Big(s\sum_{\mathcal{F}^{s,i\to N}_{[n,N,\ast]}}w(\mathcal{F})+\sum_la_{nl}(s)\sum_{\mathcal{F}^{s,i\to N,l\not\to n}_{[n,N,\ast]}}w(\mathcal{F})\Big),
    \\ B^{(i)}_N(s)&=\sum_lb_{nl}(s)\Big(\sum_{\mathcal{F}^{s,i\to N,l\to N}_{[n,N,\ast]}}w(\mathcal{F})+\sum_{\mathcal{F}^{s,i\to N,l\to \ast}_{[n,N,\ast]}}w(\mathcal{F})\Big).
\end{align}
Hence, the susceptibility for the first-passage densities is given by
\begin{align}
    \tilde{\xi}_{ij}(s)=\frac{A_N^{(i)}(s)D(s)-C(s)B_N^{(i)}(s)}{A_N ^{(j)}(s)D(s)-C(s)B_N^{(j)}(s)}.
\end{align}

\bibliography{supplement_refs.bib}